\documentstyle[aps,twocolumn,epsfig,floats]{revtex}

\newcommand{\beq}{\begin{equation}}
\newcommand{\eeq}{\end{equation}}
\newcommand{\bdis}{\begin{displaymath}}
\newcommand{\edis}{\end{displaymath}}
\newcommand{\bea}{\begin{eqnarray}}
\newcommand{\eea}{\end{eqnarray}}
\newcommand{\barr}{\begin{array}}
\newcommand{\earr}{\end{array}}
\newcommand{\boldm}[1]{\mbox{\boldmath ${#1}$}}
\newcommand{\equ}[1]{(\protect\ref{#1})}

\begin{document}

\wideabs{
\title{Dipolar interactions induced order in assemblies of magnetic
  particles} 
 
\author{R. Pastor-Satorras}

\address{The Abdus Salam International Centre 
  for Theoretical Physics (ICTP)\\
  Condensed Matter Section\\
  P.O. Box 586, 34100 Trieste, Italy}

\author{J. M. Rub{\'\i}}

\address{Departament de F{\'\i}sica Fonamental, Facultat de F{\'\i}sica,
  Universitat de Barcelona\\
  Diagonal 647, 08022 Barcelona, Spain}
\maketitle
\begin{abstract}
  We discuss the appareance of ordered structures in assemblies of
  magnetic particles. The phenomenon occurs when dipolar interactions
  and the thermal motion of the particles compete, and is mediated by
  screening and excluded volume effects. It is observed irrespective
  of the dimensionality of the system and the resulting structures,
  which may be regular or fractal, indicate that new ordered phases
  may emerge in these systems when dipolar interactions play a
  significant role.
\end{abstract}

\date{\today}
}
\newpage


\section{Introduction}

The collective behavior of assemblies of dipolar magnetic particles
is governed by the influence of several competing mechanisms and
presents a very rich and complex phenomenology \cite{rubi99} whose
ultimate interest is the identification of the emerging structures and
the subsequent analysis of its implications in the macroscopic
properties of the system. When the particles are dispersed in an
aqueous, magnetically inactive, phase, the formation of structures
results---as occurs in other soft-condensed matter systems---from the
interplay of dipolar interactions, excluded volume effects, and
thermal motion. Unlike assemblies of non-magnetic particles, they may
exhibit ordered phases emerging when dipolar interactions are
dominant, and disordered phases when those interactions are inhibited
by Brownian motion. In the cases of electro- and magneto-rheological
dispersions, the particles are heavy enough to impede Brownian
effects, and therefore chaining is the most likely event. Contrarily,
for finer magnetic particles, as monodomain particles or ferrofluids,
thermal motion becomes more important, opening the possibility to the
analysis of the interplay with dipolar interactions, and the
subsequent formation of structures. Disconnected results along this
lines have been previously reported for different systems
\cite{pastor95,pastor98a}.

Our purpose in this paper is to discuss an apparent general phenomenon
occurring in these systems: the emergence of order when magnetic
interactions and thermal motion compete. Ordered and disordered states
would then be accessible upon variation of temperature, and
consequently would correspond to different structural phases of the
system.

The paper is distributed as follows. In Section II we study the
problem of aggregation of dipolar particles in $d=2$ and $d=3$,
whereas in Sec. III we focus on a problem of different nature, namely
the adsorption of particles onto a substrate of dimension $d=1$ and
$d=2$. Finally, in the last Section we discuss the common features of
the transition to more ordered structures when dipolar interactions
play a significant role.

\section{Cluster aggregation with dipolar interactions}

The first system we consider is the cluster aggregation of particles
that experience dipolar interactions. Cluster aggregation, and in
general fractal growth phenomena \cite{vicsek92,meakin98}, are very
active fields, specially in which regards the physics of colloids. In
the context of computer models, the most notable are the
diffusion-limited aggregation (DLA) model \cite{witten81} and the
cluster-cluster aggregation model \cite{meakin83,kolb83}, which
effectively describe the fractal structure of real colloidal
aggregates \cite{weitz84}.  In this Section, we describe the extension
of the standard DLA model taking into account fully anisotropic
dipolar interparticle interactions.

\subsection{Description of the algorithm}
\label{two}

The general framework of the models we are considering is the {\em
  sequential} addition of dipolar particles of diameter $a$, with a
moment $\boldm{\mu}$ rigidly attached, to an assembly ${\cal S}$ of
rigid dipoles located at fixed positions. The incoming particles
perform a random walk under the influence of the long-range
interactions exerted by the dipoles in ${\cal S}$.  The effect of
these interactions is encoded in the following procedure: Consider
that, at a certain time step, the structure ${\cal S}$ is formed by
$N$ rigid dipoles $\boldm{\mu}_i$, located at positions $\boldm{R}_i$.
The energy of an incoming particle of magnetic moment $\boldm{\mu}$ at
position $\boldm{r}$, is then given by \cite{jackson}
\begin{equation}
  {\cal E}= \sum_{i=1}^N \frac{1}{\xi_i^3} \left\{ \boldm{\mu}
    \cdot \boldm{\mu}_i - 3 \frac{(\boldm{\mu}
    \cdot  \boldm{\xi}_i) (\boldm{\mu}_i \cdot
    \boldm{\xi}_i)}{\xi^2_i} \right\},
\label{energy}
\end{equation}
where $\boldm{\xi}_i = \boldm{r} - \boldm{R}_i$ is the relative distance
between particles. In the next time step, the dipole is moved to a new
position $\boldm{r}^*=\boldm{r} + \delta \boldm{r}$, where $\delta \boldm{r}$ is
a small increment oriented along a randomly chosen direction. The
energy in the new point $\boldm{r}^*$ is ${\cal E}^*$, and the total
variation in the energy due to the movement is $\Delta {\cal E} = {\cal
  E}^* - {\cal E}$. If $\Delta {\cal E}<0$, the movement is accepted with
probability $1$; otherwise, it is accepted with probability $\exp(-\Delta
{\cal E}a^3/\mu^2 T_r)$, where $T_r=a^3 k_{\mbox{\rm\scriptsize B}} T /
\mu^2$ is a dimensionless reduced temperature, relating the intensity of
the dipolar interactions $\mu$ to the actual temperature $T$ of the
system. In case of not being accepted, the trial position
$\boldm{r}^*$ is discarded and a new one generated.

With the previous algorithm, we have taken into account the
translations dynamics of the particles. For the rotational dynamics of
the dipoles, we apply the following prescription: The initial
orientation of the dipoles $\boldm{\mu}^0$ is assigned at random. After
every accepted movement, the moment of the random walker is oriented
along the direction of the total magnetic force on its position.
After the dipole becomes finally attached to the assembly, its moment
undergoes a final relaxation, and its direction does not change any
more. Indeed, this fact assumes that the relaxation time for the
orientation of a particle is very short compared with the
characteristic time scale of the movement of its center of mass.

The simulation of the particle-cluster aggregation process starts with
a particle (the seed) located at the origin of coordinates, and
bearing a randomly oriented moment $\boldm{\mu}_0$. The following
particles are released from a random position on a sphere of radius
${\cal R}_{\rm in}$ centered on the seed. The particles have an
initial moment $\boldm{\mu}_i$ assigned at random. Each particle
undergoes a Brownian motion, according to the given prescription,
until it either contacts the cluster or moves away a distance larger
than ${\cal R}_{\rm out}$. In this case, the particle is removed and a
new one is launched. In the present simulations, we have used the
values ${\cal R}_{\rm out}= 2 \bar{{\cal R}}$ and ${\cal R}_{\rm
  in}={\cal R}_{\rm out} - 5a$, where $\bar{{\cal R}}$ is the radius
of the cluster, measured in units of the particle diameter $a$.  The
particles stick to the cluster when they overlap one or more particles
previously aggregated. The new dipole is then attached in the point of
its last trajectory where it is first tangent to the surface of the
cluster.

\subsection{Numerical results in $d=3$}
\label{one}

Simulations of cluster aggregation of dipolar particles in $d=3$
are extremely time-consuming. Therefore, we report only results for
the limits $T_r=0$ and $T_r=\infty$. The case $T_r=0$ corresponds to
infinitely strong dipolar interactions; the steps during the random
walk are only accepted if they lower the total dipolar energy. For
$T_r=\infty$ all movements are accepted and we must recover the standard
DLA model \cite{witten81}

\begin{figure}[t]
  \centerline{\epsfig{file=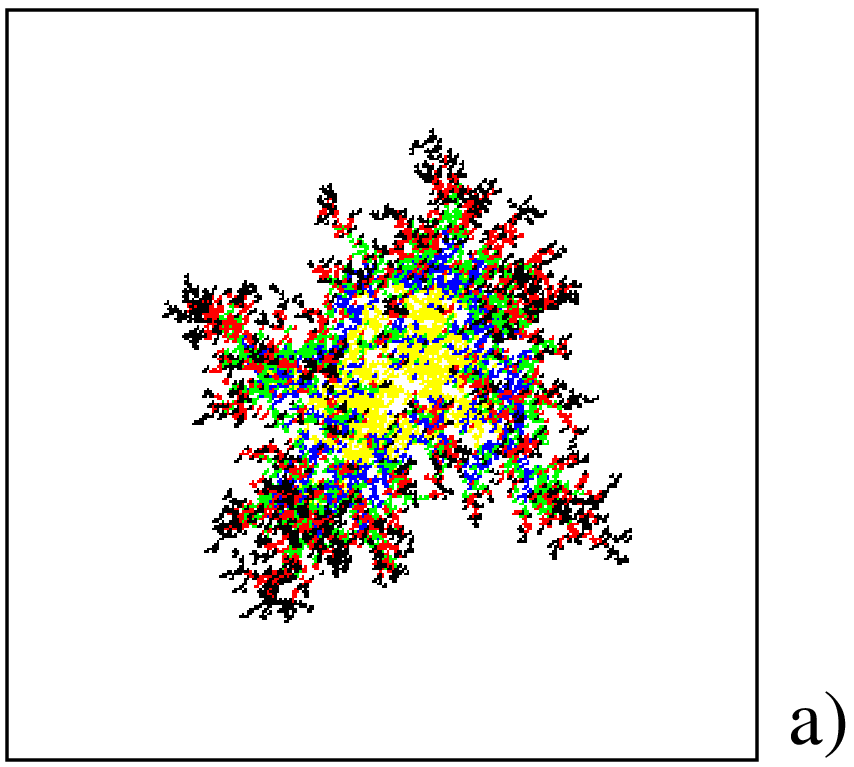, width=7cm}}
  \centerline{\epsfig{file=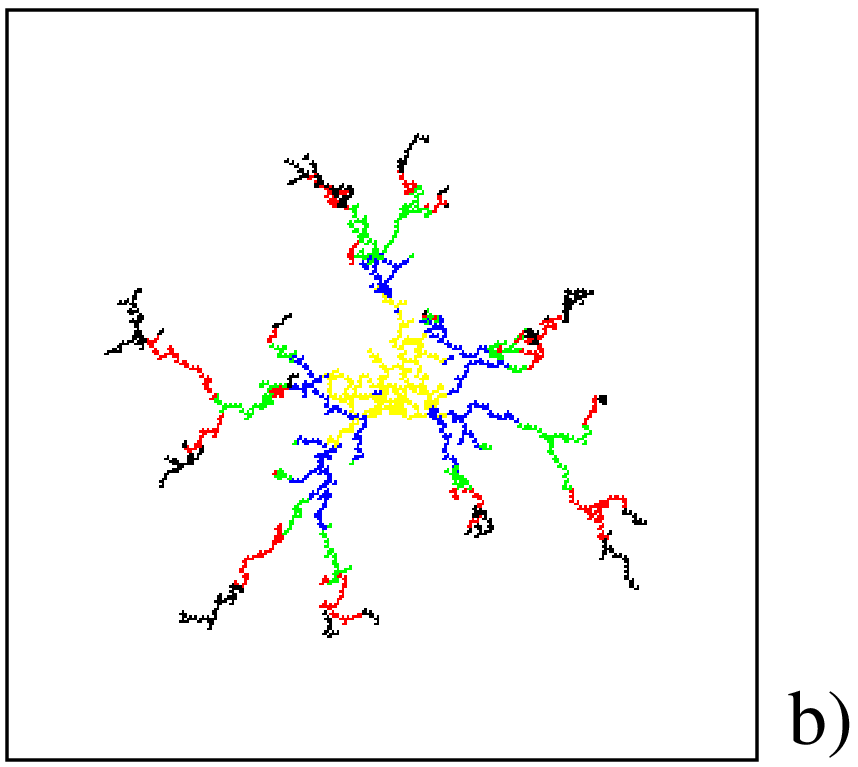, width=7cm}}
  \vspace*{0.25cm}
  \caption{Projection onto the $x-y$ plane of two typical dipolar
    clusters in $d=3$, grown on a cubic box of size $L=150a$. (a)
    $T_r=\infty$, $D_f=2.47\pm0.03$. (b) $T_r=0$, $D_f=1.37\pm0.03$ (see
    text).}
  \label{fig:dla3d}
\end{figure}

In Figure~\ref{fig:dla3d} we show two typical clusters grown at
$T_r=0$ and $T_r=\infty$ in a box of size $L=150a$. The growth was stopped
when the radius of the clusters reached a predetermined distance to
the boundary of the box. In the case $T_r=0$, the cluster was composed
by $N\sim1500$, whereas for $T_r=\infty$ we have $N\sim10000$. A first conclusion
can be immediately drawn: At high temperatures, the clusters are more
compact, having an roughly spherical shape.  When decreasing the
temperature, the average density of the clusters diminishes; at
$T_r=0$, the clusters have an essentially digitated structure of very
low density.

A more quantitative estimate of this concept can be obtained by
measuring the fractal dimension $D_f$ of the clusters
\cite{vicsek92}. The value of $D_f$ is determined from a log-log plot
of the radius of gyration $R_g(N)$ as a function of the number of
particles in the cluster $N$, using the relationship \cite{vicsek92}
\begin{equation}
  R_g(N) = \left(\frac{1}{N} \sum_{i=1}^N R_i^2 - \frac{(\sum_{i=1}^N
      R_i)^2}{N^2} \right)^{1/2} \simeq N^{1/D_f}.
\end{equation}
The value computed for the aggregation at $T_r=\infty$ is $D_f=2.47\pm0.03$,
in good agreement, within error bars, with the best estimate for DLA
in $d=3$, namely $D_f^{\rm DLA}=2.495\pm0.005$ \cite{tolman89}. On the
other hand, for the simulations at zero temperature, we obtain a much
smaller value, $D_f=1.37\pm0.03$.

These strong variations in the structure of the cluster with
temperature can be understood through a very simple and intuitive
argument. The aggregation process depends on the competition between
dipolar forces and thermal agitation. Dipolar interactions favor the
formation of chains of aligned dipoles, thus minimizing the dipolar
energy, while the thermal motion tends to randomize the process and
produce a higher degree of ramification. This branching process is
mediated by the {\em screening} of the inner regions of the cluster by
the most external branches. The chances for a random walker to surpass
the external active region and reach the core of the cluster are very
small, and this effect induces a high branching ratio and a large
fractal dimension.  The balance between these two effects, branching
and growth, can be measured by estimating the relative probabilities
of growth, $p_g$, of an already existing branch and its splitting,
$p_s$, creating thus a new branch. Assuming that the branches are
composed by relaxed (parallel) dipoles, pointing along the axis of the
branch, the addition of a new particle on the tip of an existing
branch will increase the energy of the aggregate by an amount
$\Delta{\cal E}_g = -2 \mu^2/a^3$ (we only consider the interaction of
the new particle with its nearest neighbor). We can associate to this
growth event a relative probability given by a Boltzmann factor, and
set $p_g=\exp(-\Delta{\cal E}_g a^3/\mu^2 T_r)=\exp(2/T_r)$.  On the
other hand, in order to minimize the energy, an incoming particle,
inducing a branching event, will set its dipole anti-parallel to its
nearest-neighbor in the existing branch, thus increasing energy by
$\Delta{\cal E}_s = - \mu^2/a^3$. The corresponding splitting
probability is then $p_s =\exp(1/T_r)$. We have therefore
\begin{equation}
  \frac{p_g}{p_s}=\exp(1/T_r).
\end{equation}
At high temperature, $p_g\simeq p_s$; the growth and splitting of a given
branch are equally likely events. The effect of the dipolar
interactions is overcome by the thermal disorder and we must recover
the original DLA with no interactions. At low temperatures, the
splitting probability is vanishingly small compared with $p_g$, and one
should expect to observe in this limit very fingered aggregates, with
little branching, and a fractal dimension closer to $1$, as is indeed
the case in Fig.~\ref{fig:dla3d}.

Considerations about the fractal dimension allows also to draw
conclusions on the macroscopic aggregation of individual clusters. At
low concentrations of dipolar particles, it is reasonable to assume
that particles first form clusters, which on their hand coalesce later
to form larger structures. If the mutual interactions between clusters
are negligible (due to screening of the interactions in a large
reservoir of clusters with random total moment) the coalescence of
clusters is mediated by Brownian motion. At low temperatures, the
cluster dimension is low. Let $D_f^{(1)}$ and $D_f^{(2)}$ be the
fractal dimension of any two clusters in the reservoir (of course,
$D_f^{(1)}=D_f^{(2)}$). Since $D_f^{(1)} + D_f^{(2)} <3$, the clusters
are transparent, have little mutual interactions and behave as
essentially independent objects \cite{witten98}. At high temperatures,
however, the fractal dimension is larger and fulfills the condition of
mutual opacity $D_f^{(1)} + D_f^{(2)} >3$ \cite{witten98}. Assuming
that $D_f$ is an increasing function of $T_r$ (as is the case in
$d=2$, see next section), we predict the existence of a critical
temperature $T_r^c$ for which $D_f^{(1)} + D_f^{(2)}=3$, setting the
threshold above which cluster coalescence is possible, leading to
the formation more complex structures.

\begin{figure}[t]
  \centerline{\epsfig{file=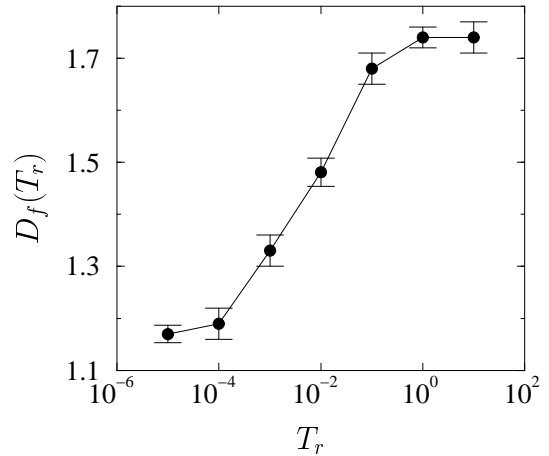, width=7cm}}
  \vspace*{0.25cm}
  \caption{Fractal dimension $D_f$ for dipolar clusters in $d=2$ as a
    function of the reduced temperature $T_r$.}
  \label{fig:dladimen}
\end{figure}

\subsection{Numerical results in $d=2$}

Simulations in $d=2$ are much less costly, and afford the possibility
of truly exploring the effect of temperature on the cluster structure.
Moreover, simulations on a plane are also realistic and interesting on
their own, because of their relevance in understanding phase
transitions in Langmuir monolayers \cite{kaganer99}. Far from
equilibrium, the condensed phase grows at the expense of the liquid
phase, forming clusters of different shapes. The phospholipids that
make up the Langmuir monolayer experience repulsive dipolar
interactions, which must play a major role in determining the
morphology of the condensed aggregates.

In Figure~\ref{fig:dladimen} we plot the fractal dimension $D_f$ for
dipolar clusters in $d=2$ as a function of the reduced temperature
$T_r$. This figure confirms the arguments given in Sec.~\ref{one}. At
high temperatures the fractal dimension is large ($D_f=1.74\pm0.03$ at
$T_r=10$) and compatible with the best estimates for free DLA
($D_f^{\rm DLA}=1.715\pm0.004$, \cite{tolman89}). At low temperatures,
on the other hand, the fractal dimension is much smaller, and closer
to $1$ ($D_f=1.13\pm0.01$ at $T_r=0$). That is, the lower the
temperature, the stronger the screening effects of the branches,
leading to an essentially digitated structure composed by fingers with
very few side branches. At this respect, a significant difference with
the case $d=3$ is that now clusters are always mutually opaque,
$D_f^{(1)} + D_f^{(2)} >2$, for all values of $T_r$.

\begin{figure}[t]
  \centerline{\epsfig{file=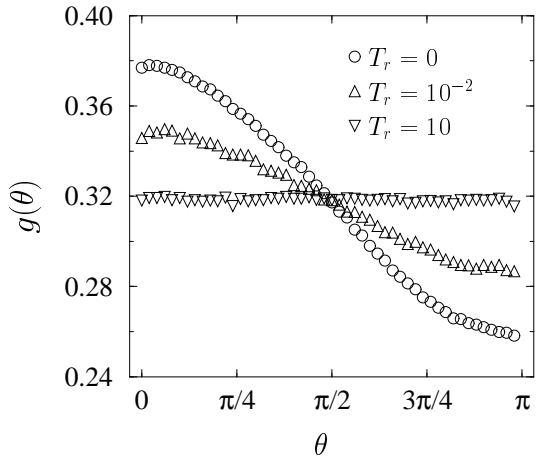, width=7cm}}
  \vspace*{0.25cm}
  \caption{Angular correlation function $g(\theta)$ in dipolar clusters in
    $d=2$ for different values of the reduced temperature $T_r$.}
  \label{fig:dlacorrels}
\end{figure}

Another remarkable feature of this system is the {\em ordering}
induced by the dipolar interactions. A convenient way to measure this
order is the angular correlation function $g(\theta)$ of the dipoles
composing the cluster, defined as the fraction of all pairs of
dipoles, irrespective of the position, forming a relative angle
$\theta$. This correlation function is shown in
Fig.~\ref{fig:dlacorrels} for different temperatures. At high
temperatures we observe that the system is completely disordered, with
all orientations between dipoles being equally probable.  More
interestingly, at low temperatures we observe the emergence of an
ordered phase, in which dipoles are parallel with high probability,
while suppressing the anti-parallel (more energy costly)
configurations. That is, there is an order-disorder transition
(reflected in the angular correlation function $g$) between an ordered
phase at the low-temperature limit (in the absence of any thermal
disorder) in which the orientation of the dipoles is strongly
correlated, and a disordered phase in the high-temperature limit, in
which dipoles show an uncorrelated behavior, imposed by the intrinsic
fractal geometry of the clusters.

The structures that we have shown in this section have been observed
experimentally \cite{indivery96,lefebure98}. In the latter
experiments, ferromagnetic particles of different sizes are
transferred to a Langmuir monolayer at different pressures. The main
observed feature is that the larger the diameter of the particles, the
more digitated the emerging cluster. These experimental results can be
interpreted by using our argument which is able to predict the shape
of the clusters in terms of the reduced temperature. That equation
indicates that, under those circumstances, the probability of a
cluster to grow is much higher than the probability for splitting.
Conversely, when the diameter of the particles is small, both
probabilities equilibrate, resulting in denser circular compact
aggregates, also in accordance with experimental results.

\section{Chaining of dipolar particles onto a substrate}

In this Section we present results for a model of adsorption in which
the adsorbing particles experience dipolar interactions. The
irreversible adsorption of colloidal particles onto a solid surface is
a very relevant subject, with plenty of practical and technological
applications \cite{bartelt91,evans93}. The basic models proposed so
far in an attempt to understand the physics of adsorption are the
random sequential adsorption  model \cite{renyi63} and the
ballistic model (BM) \cite{talbot92}. The BM model is a good
approximation for the adsorption of large colloidal particles. In this
model, particles arrive at the surface following straight vertical
trajectories. When an incoming particle fails to reach the substrate
directly, it is allowed to roll down over the previously adsorbed
ones, until it reaches an equilibrium position. Particles eventually
resting on the surface are adsorbed. Otherwise, they are rejected.

These models, and their subsequent generalizations, consider only very
short range---usually hard-core--- interactions. Here we will
illustrate the profound effect of interparticle dipolar interactions
in the adsorbed phase, for the case of ballistic adsorption.

\subsection{Description of the algorithm}

We consider the irreversible adsorption of dipolar particles that
diffuse in a semi-infinite $d+1$-dimensional space and adsorb upon
contact onto a boundary $d$-dimensional hyperplane (the substrate). In
this case, the assembly of particles ${\cal S}$ under consideration is
the phase formed by all the particles adsorbed at given time, located
at positions $\boldm{R}_i$ and carrying a rigid moment $\boldm{\mu}_i$.
Since we are considering an irreversible process, the adsorbed
particles are fixed and cannot diffuse on top of the substrate.  The
algorithm used to simulated the random walk of the incoming particles
is exactly as described in section \ref{two}, differing of course in
the boundary conditions.

Simulations start with a clean substrate. Particles are released
sequentially at random positions over a certain initial height $h_{\rm
  in}$ above the substrate, and with an initial random orientation
$\boldm{\mu}_0$. The particles perform a random walk until they either
hit the substrate and become adsorbed or move away a distance larger
than $h_{\rm out}$.  In this case, the particle is removed and a new
one released.  We have chosen the parameters $h_{\rm in}=15a$ and
$h_{\rm out}=20a$, which are reasonable given the power-law decay of
dipolar interactions \equ{energy}. The rotational dynamics of the
dipoles is the same as in the case of cluster aggregation, namely,
dipoles relaxing at every accepted step.  When a particle reaches a
position very close to the surface (less than one particle diameter
$a$), it is attached to the substrate following the BM rules. This
procedure assumes therefore the presence of a strong short-ranged
interaction between substrate and particles, which overcomes dipolar
interactions at very short distances.  The newly adsorbed particles
experience one last relaxation, and their moment does not change any
more afterwards.  The adsorbing surface has linear dimension $L$ and
we impose periodic boundary conditions in all directions but in the
perpendicular to the plane. The system sizes considered in this work
range between $L=150a$ and $L=250a$.

\begin{figure}[t]
  \centerline{\epsfig{file=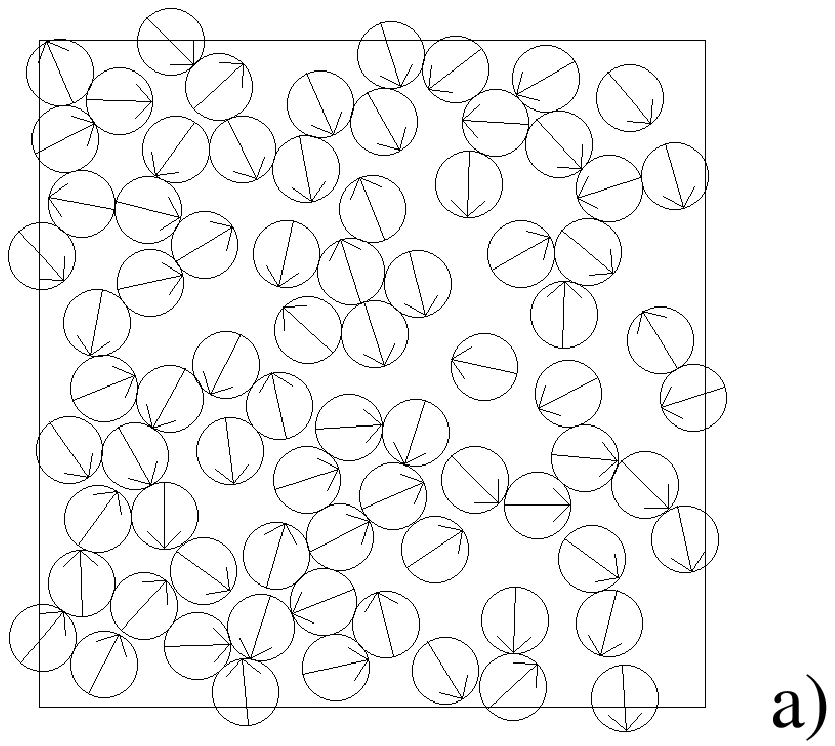, width=7cm}}
  \centerline{\epsfig{file=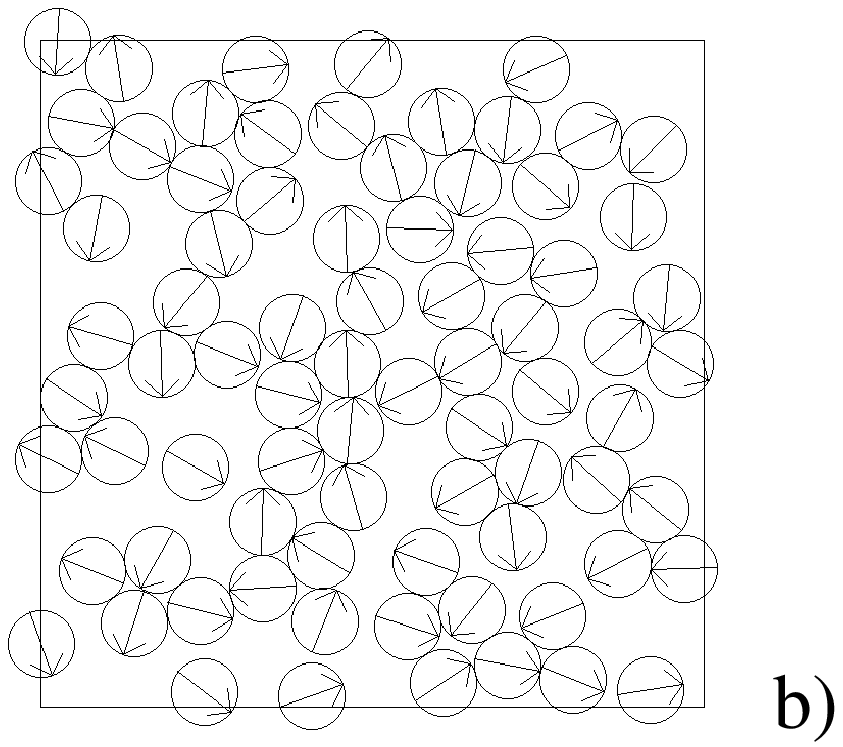, width=7cm}}
  \vspace*{0.25cm}
  \caption{Saturated configurations in $d=2$, system size $L=10a$.
    (a) Free BM, $T_r=\infty$. (b) With dipolar interactions, at $T_r=0$.
    The particles extend out of the box to show the effect of periodic
    boundary conditions.}
  \label{fig:bm3d}
\end{figure}

\subsection{Numerical results in $d=2$}

Again, numerical simulations in $d=2$ are very time-consuming and
permit only to consider very small system sizes. In
Figure~\ref{fig:bm3d} we have represented two typical saturated
configurations, corresponding to the adsorption according to the rules
of standard BM ($T_r=\infty$) and in the presence of dipolar interactions
($T_r=0$). Inspection of these figures allows to conclude that the
packing of the substrate is higher in presence of interactions. A
numerical measure of the packing is given by the {\em jamming limit}
$\theta_\infty$, defined as the maximum fraction of the surface covered by
particles at the saturation limit. Our simulations provide a value
$\theta_\infty=0.622\pm0.003$ for the dipolar adsorption at $T_r=0$, whereas the
standard BM ($T_r=\infty$) corresponds to $\theta_\infty^{\rm BM}=0.61048\pm0.00013$
\cite{thompson92}. This increase in the packing of the saturated
surface can be traced back to the attractive interactions between the
adsorbed phase and the incoming particles. In the case of standard
free BM, the structure of the adsorbed phase is due exclusively to
excluded volume effects: the already attached particles impede the
adsorption of new particles at random positions, and allow only the
occupation of empty spaces. This is essentially a random, disordering,
effect. The presence of dipolar interactions, attracting the random
walker to the surface, diminishes the excluded volume effects and
leads to configurations of higher packing.

\begin{figure}[t]
  \centerline{\epsfig{file=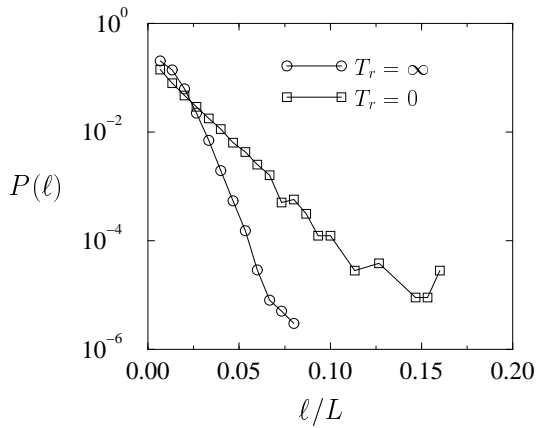, width=7cm}}
  \vspace*{0.25cm}
  \caption{Chain-length density function $P(\ell)$ for different values
    of the reduced temperature $T_r$, for dipolar adsorption in
    $d=1$. System size $L=250a$.}
  \label{fig:bmlength}
\end{figure}

\subsection{Numerical results in $d=1$}

By restricting the adsorption onto a line, we can explore more
thoroughly the interplay between interactions and the disordering
excluded volume effects. As we can conclude from examination of
Fig.~\ref{fig:bm3d}, the reason of the higher coverage is the larger
tendency of dipolar particles to form connected structures at low
temperatures. In $d=1$ these connected structures are identified with
chains. We therefore propose as a measure of order the chain-length
density function $P(\ell)$, defined as the average number of chains of
length $\ell a$, per unit length of substrate. From this function we can
compute the remaining interesting properties of the adsorbed phase,
like the jamming limit
\begin{equation}
  \theta_\infty = \sum_{\ell} \ell P(\ell).
\end{equation}

In Figure~\ref{fig:bmlength} we have represented the chain-length
density function for dipolar adsorption at $T_r=0$ and $T_r=\infty$
(free BM). At high temperature we observe a very fast
(super-exponential) decay of $P(\ell)$, indicative of a lack of any
characteristic length; disorder has suppressed the formation of
correlated chains in this case. At low temperatures, however,
$P(\ell)$ decays exponentially, $P(\ell)\sim\exp(-\ell/\xi)$, with a
well-defined correlation length (or average chain size)
$\xi=4.73\pm0.02$. Chains have now the possibility of growing very
large, and a single chain is seen to cover more than $15\%$ of the
whole substrate.

\begin{figure}[t]
  \centerline{\epsfig{file=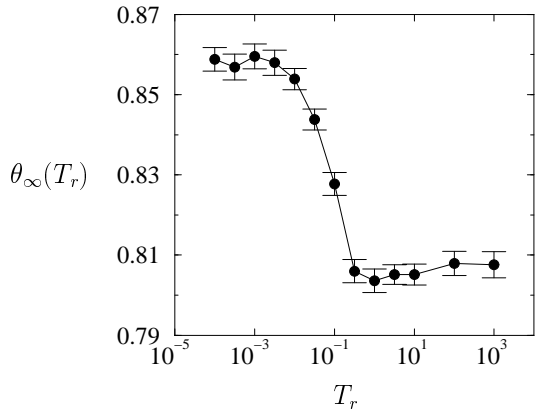, width=7cm}}
  \vspace*{0.25cm}
  \caption{Jamming limit $\theta_\infty$ as a function of the reduced
    temperature $T_r$, for dipolar adsorption in $d=1$.}
  \label{fig:bmjam}
\end{figure}

Finally, we have plotted in Fig.~\ref{fig:bmjam} the jamming limit as
a function of the reduced temperature $T_r$. At high temperature
($T_r=10$) we recover the theoretical value $\theta_\infty^{\rm
  BM}=0.8079\pm0.0003$ \cite{talbot92}. On the other hand, at low
temperatures the higher packing due to the chaining of the particles
leads to a higher coverage, namely $\theta_\infty=0.855\pm0.002$, which
differs from the free case in about  $6\%$, a notable variation in
the context of adsorption phenomena. Note, moreover, the strong
resemblance of Figs.~\ref{fig:dladimen} and \ref{fig:bmjam}, which
depict analogous magnitudes in both models presented.

\section{Conclusions}

The results presented in this paper render the following
conclusions. The overall behavior of assemblies of magnetic particles 
is the result of the competition of factors of different nature:
dipolar interactions, thermal disorder (Brownian motion) and screening 
(for aggregation) or excluded volume effects (for adsorption). 

The concept of reduced temperature enables us to ascertain the global
form of the clusters. A more precise analysis of the geometry of the
structures, able to account for their local structure, needs however
the use of fractal concepts, and in particular of fractal
dimension. In the cases we have discussed, magnetic particles may
assemble basically into structures of effective $D_f=1$, corresponding
to the chaining process on a substrate mediated by excluded volume
interactions, or into structures with values of $D_f$ comprised in the
interval $]1,2[$, ranging from digitated to spherical clusters. 

In the case of dipolar cluster aggregation, we have observed the
existence of transitions between an ordered, or {\em quasi-ordered},
phase emerging at low values of the reduced temperature (and therefore
comprising the low temperature and large particle size cases) and a
disordered phase for high values of the reduced temperature. The low
temperature phase is characterized by the existence of long-range
correlations between dipoles. In the $d=2$ case, our predicted results
have been corroborated in experiments of aggregation performed on
Langmuir monolayers. This technique has revealed to be a very useful
tool in the characterization of the resulting two-dimensional
structures.

Finally, it is worth pointing out that the existence of order at the
mesoscopic level may have implications in the behavior of macroscopic 
quantities, such as the magnetization of the system.

\acknowledgements

\section*{}

The work of R.P.S. has been supported by the European Network under
Contract No.  ERBFMRXCT980183. J.M.R. acknowledges financial support
by CICyT (Spain), Grant No. PB98-1258, and by the INCO-COPERNICUS
programme of the European Commission under Contract No.
IC15-CT96-0719.


\begin{thebibliography}{10}

\bibitem{rubi99}
J.~M. Rub{\'\i} and J.~M.~G. Vilar, {J.} Phys. C (in press).

\bibitem{pastor95}
R. Pastor-Satorras and J.~M. Rub\'{\i}, Phys. Rev. E {\bf 51},  5994  (1995).

\bibitem{pastor98a}
R. Pastor-Satorras and J.~M. Rub\'{\i}, Phys. Rev. Lett. {\bf 80},  5373
  (1998).

\bibitem{vicsek92}
T. Vicsek, {\em Fractal Growth Phenomena}, 2nd ed. (World Scientific,
  Singapore, 1992).

\bibitem{meakin98}
P. Meakin, {\em Fractals, Scaling and Growth far from Equilibrium} (Cambridge
  University Press, Cambridge, 1998).

\bibitem{witten81}
T.~A. Witten and L.~M. Sander, Phys. Rev. Lett. {\bf 47},  1400  (1981).

\bibitem{meakin83}
P. Meakin, Phys. Rev. Lett. {\bf 51},  1119  (1983).

\bibitem{kolb83}
K. Kolb, R. Botet, and R. Jullien, Phys. Rev. Lett. {\bf 51},  1123  (1983).

\bibitem{weitz84}
D.~A. Weitz and M. Oliveria, Phys. Rev. Lett. {\bf 52},  1433  (1984).

\bibitem{jackson}
J.~D. Jackson, {\em Classical Electrodynamics}, 2nd ed. (John Wiley \& Sons,
  New York, 1975).

\bibitem{tolman89}
S. Tolman and P. Meakin, Phys. Rev. A {\bf 40},  428  (1989).

\bibitem{witten98}
T.~A. Witten, Rev. Mod. Phys. {\bf 70},  1531  (1998).

\bibitem{kaganer99}
V.~M. Kaganer, H. M{\"o}hwald, and P. Dutta, Rev. Mod. Phys. {\bf 71},  779
  (1999).
  
\bibitem{indivery96} 
G. Indivery, A. C. Levi, A. Gliozzi, E. Scalas,
  and H. M{\"o}wald, Thin Solid Films {\bf 284--285}, 106 (1996).

\bibitem{lefebure98}
S. Lefebure, C. Menager, V. Cabuil, M. Assenheimer,
F.  Gallet, and C. Flament, J. Phys. Chem. B {\bf 102},  2733  (1998).

\bibitem{bartelt91}
M.~C. Bartelt and V. Privman, Int. J. Mod. Phys. B {\bf 5},  2883  (1991).

\bibitem{evans93}
J. Evans, Rev. Mod. Phys. {\bf 65},  1281  (1993).

\bibitem{renyi63}
A. R\'{e}nyi, Sel. Trans. Math. Stat. Prob. {\bf 4},  203  (1963).

\bibitem{talbot92}
J. Talbot and S.~M. Ricci, Phys. Rev. Lett. {\bf 68},  958  (1992).

\bibitem{thompson92}
A.~P. Thompson and E.~D. Glandt, Phys. Rev. A {\bf 46},  4639  (1992).

\end{thebibliography}
\end{document}